\documentclass[11pt]{article}

\pdfoutput=1 

\usepackage{jheppub}

\usepackage{graphicx}
\usepackage{color}
\usepackage{amsmath,amssymb,amsfonts}
\usepackage{verbatim}   
\usepackage{subfigure}  
\usepackage{acronym}
\usepackage{hyperref}
\usepackage{mathrsfs}
\usepackage{slashed}
\usepackage{multirow}
 \usepackage{cancel}
 \usepackage{url}
 \usepackage{hyperref}
 
\newcommand{\be}{\begin{equation}} 
\newcommand{\ee}{\end{equation}} 

\def\beq{\begin{equation}\begin{aligned}}
\def\eeq{\end{aligned}\end{equation}}

\def\OO{\mathcal{O}}
\def\dd{{\rm d}}

\title{UltraViolet Freeze-in}

\author{Fatemeh Elahi,  Christopher Kolda, and James Unwin}

\affiliation{Department of Physics, University of Notre Dame,\\ 225 Nieuwland Science Hall,  Notre Dame, IN 46556}

\abstract{
If dark matter is thermally decoupled from the visible sector, the observed relic density can potentially be obtained via freeze-in production of dark matter. Typically in such models it is assumed that the dark matter is connected to the thermal bath through feeble renormalisable interactions. Here, rather, we consider the case in which the hidden and visible sectors are coupled only via non-renormalisable operators. This is arguably a more generic realisation of the dark matter freeze-in scenario, as it does not require the introduction of diminutive renormalisable couplings. We examine general aspects of freeze-in via non-renormalisable operators in a number of toy models and present several motivated implementations in the context of Beyond the Standard Model (BSM) physics. Specifically, we study models related to the Peccei-Quinn mechanism and $Z'$ portals.}

\begin{document}

\hfill \vspace{-5mm} 22nd October 2014

\maketitle


\section{Introduction}

\begin{figure}[t!]
\begin{center}
\includegraphics[height=55mm]{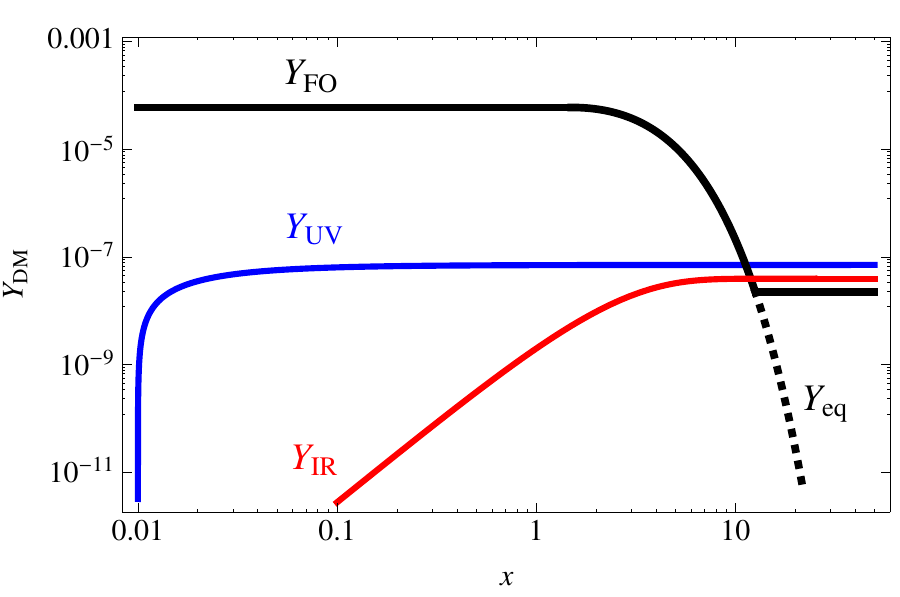}
\vspace{-2mm}
\caption{A schematic plot of the evolution of the DM yield $Y\equiv n_{\rm DM}/S$ with respect to inverse scaled temperature $x\propto T^{-1}$ for the freeze-out $Y_{\rm FO}$, IR freeze-in $Y_{\rm IR}$ and UV-freeze-in $Y_{\rm UV}$ scenarios.
\label{Fig1} \vspace{-5mm}}
\end{center}
\end{figure}

In the {\em freeze-out} paradigm of dark matter (DM), such as the much studied WIMP scenario, the DM is initially in thermal equilibrium and its abundance evolves with its equilibrium distribution until it decouples from the thermal bath. After decoupling the DM comoving number density is constant and (for appropriate parameter values) can give the observed relic density. Models of {\em freeze-in} DM \cite{Hall:2009bx,Cheung:2010,Blennow:2013jba,Kolda:2014ppa,axion,FI,Asaka,3/2,McDonald:2001vt,McDonald:2008ua,Harling:2008px,Mambrini} provide a very different picture of the evolution of the DM abundance. In this setting it is supposed that the DM number density is initially negligible but over time an abundance suitable to match the relic density is produced due to interactions in the thermal bath involving a suppressed portal operator. This is illustrated in Fig.~\ref{Fig1}. For the DM abundance to be initially negligible, and subsequently set by the freeze-in mechanism (rather than freeze-out), the hidden sector must be thermally decoupled from the visible sector bath at all times, which implies that the portal operators must be extremely small. For instance, for TeV scale DM produced via $2\rightarrow2$ scattering of bath states involving a renormalisable portal interaction the coupling dressing this operator should be typically $\lesssim10^{-7}$ in order to avoid equilibration of the DM with the visible sector \cite{Cheung:2010}. Such DM states are sometimes referred to as {\em feebly interacting massive particles}, or FIMPs.

Freeze-in, as a general mechanism for DM production, was proposed only recently \cite{Hall:2009bx}\footnote{This framework builds upon earlier specific realisations, most notably the production of right-handed neutrinos \cite{Asaka}, axinos \cite{axion}, and gravitinos \cite{3/2}, and see also \cite{McDonald:2001vt,McDonald:2008ua,Harling:2008px}.} and thus many important aspects remain to be studied. In particular, a huge class of models has been largely neglected and the purpose of this paper is to rectify this. Freeze-in using renormalisable interactions has been considered in some detail  \cite{Hall:2009bx,Cheung:2010,FI}; here instead we examine the alternative possibility, that freeze-in production proceeds via non-renormalisable operators.  A suitable DM abundance can potentially  be generated by freeze-in via such effective operators, which we refer to as {\em UltraViolet (UV) freeze-in}, and in this case the DM abundance depends sensitively on the reheat temperature. Conversely, we use {\em InfraRed (IR) freeze-in} to refer to the class of models in which the sectors are connected via renormalisable operators, in which case the DM abundance is set by IR physics and is independent of the reheat temperature.  The different thermal histories associated to these DM frameworks are illustrated in Fig.~\ref{Fig1}.

The two basic premises of the general freeze-in picture are that
\hspace*{-5mm}
\begin{itemize}
\item The hidden and visible sectors are thermally disconnected,
\item The inflaton decays preferentially to the visible sector, not reheating the hidden sector.
\end{itemize}
Consequently, it is a model independent statement that, due to the out-of-equilibrium dynamics, DM production will proceed through freeze-in via any non-renormalisable operator which is not forbidden by symmetries. Further, the expectation from UV completions of the SM is that distinct sectors of the low energy theory are generically connected by UV physics.

On the other hand, IR freeze-in relies on a rather special construction in which the (renormalisable) portal operators have diminutive couplings, however the na\"ive expectation is that dimensionless parameters should be near unity. Whilst such feeble couplings are not inconceivable (the electron Yukawa $Y_e\sim10^{-6}$ is one example), such decoupling is readily achieved if the visible sector and hidden sector are only connected via high dimension operators.

The UV freeze-in scenario bears some resemblance to models of non-thermal DM \cite{Chung:1998zb}. The two frameworks both require the DM to be thermally decoupled from the visible sector, they also rely on particular realisations of inflation, and both lead to a DM abundance which is dependent on the reheat temperature. However, there are also important distinctions between these frameworks. In non-thermal DM, the DM has sufficiently small couplings with the visible sector such that energy exchange between the sectors is negligible and the DM relic density is set primarily by inflaton decay. In contrast, in UV freeze-in the DM is dominantly populated by energy transfer from the visible sector to the hidden sector.

In this paper we examine a range of motivated operators for UV freeze-in and discuss potential connections with other aspects of high scale physics. The paper is structured as follows: In Sect.~\ref{S2} we develop the physics behind UV freeze-in using a number of toy models which exemplify several interesting features.  In particular, we discuss high dimension operators with many body final states. Further, we examine examples in which a field involved in the portal operator develops a vacuum expectation value (VEV). We consider the constraints which arise from avoiding sector equilibration in Sect.~\ref{EqCo} and find that this leads to bounds on the parameter space, but that large classes of viable models can be constructed. Subsequently, in Sect.~\ref{S3} we propose a number of simple models, motivated by beyond the Standard Model  (BSM) physics, which realise the UV freeze-in picture. Specifically, we consider possible connections with  axion models and $Z'$ portals. We also comment on the prospect of identifying the scale of UV physics given the DM mass, portal operator and the magnitude of the reheat temperature. In Sect.~\ref{S4} we provide a brief summary, alongside our closing remarks.


\section{General possibilities for UV freeze-in}
\label{S2}

The possibility of freeze-in via non-renomalisable operators has been briefly discussed in \cite{Hall:2009bx,Blennow:2013jba,Harling:2008px,Kolda:2014ppa,Mambrini}. One of the distinguishing features of UV freeze-in is that  DM production is dominated by high temperatures, and so the abundance is sensitive to the reheat temperature $T_{\rm RH}$. Whilst this possibility has been previously remarked upon as less aesthetic due to the dependence on the unknown value of $T_{\rm RH}$, it is nevertheless very well motivated as it is a generic expectation that sectors which are decoupled at low energy may communicate via high dimension operators. In this section we examine some general classes of toy models in which the hidden and visible sector are connected only by effective contact operators.


\subsection{Dimension-five operators with two and three-body final states}
\label{S2.1}

We shall start by discussing the simple toy model of UV freeze-in outlined by Hall, March-Russell \& West \cite{Hall:2009bx} (see also \cite{McDonald:2008ua} for a similar model in the context of supersymmetry); this will provide a basis from which to examine more realistic scenarios in subsequent sections. In this toy model a scalar DM state $\varphi$ freezes-in due to a dimension five operator of the form\footnote{As $\varphi$ appears linearly in the operator it can not be stabilised by a simple $Z_2$. We examine the issue of stability in Sect.~\ref{S3} for specific models, but note here that an enlarged symmetry could accommodate this operator and stabilise $\varphi$. Alternatively, $\varphi$ could be an unstable hidden sector state which decays to the DM. At present we use this example as a simple toy model to illustrate freeze-in via non-renormalisable operators.} 
\beq
\mathcal{L}\supset \frac{1}{\Lambda}\varphi \bar\psi_1\psi_2\phi~,
\label{JMR}
\eeq
 where $\phi$ is a boson in the thermal bath, $\psi_i$ are bath fermions, and $\Lambda$ is the mass scale at which the effective operator is generated. Throughout this section we shall use $\varphi$ and $\chi$ to denote scalar and fermion hidden sector states,  respectively, and use $\phi$ and $\psi$ to indicate scalars and fermions in the thermal bath. Let us suppose, for the time being, that $\phi$ does not develop a VEV. An abundance of $\varphi$ can freeze-in via $2 \rightarrow 2$ scattering processes involving the bath states: $\phi \psi_1 \rightarrow  \varphi \psi_2$.

 The change in number density $n$ can be described by the Boltzmann equation (see e.g.~\cite{Kolb:1990vq})
 \beq
\dot{n}_\varphi+3H n_\varphi=
\int {\rm d} \Pi_\phi {\rm d} \Pi_{\psi_1} {\rm d} \Pi_{\psi_2}  {\rm d} \Pi_{\varphi} & (2\pi)^4\delta^{(4)}(p_{\psi_1}+p_{\phi}-p_{\psi_2}-p_\varphi)\\
&\times\Big[
|\mathcal{M}|^2_{ \phi \psi_1 \rightarrow  \varphi \psi_2}f_\phi f_{\psi_1} -|\mathcal{M}|^2_{ \varphi \psi_2 \rightarrow  \phi \psi_1}f_\varphi f_{\psi_2} 
\Big]~,
\label{BE}
\eeq 
where $\dd \Pi_i\equiv \frac{\dd^3 p_i}{(2\pi)^3} \frac{1}{2E_i}$ and $f_i$ is the distribution function for a given state. We shall assume that the various states are in thermal equilibrium and thus Maxwell-Boltzmann distributed, $f_i\sim e^{-E_i/T_i}$, with the visible sector states $\phi,\psi_i$ distributed with respect to the temperature of the  thermal bath $T$, whereas the DM $\varphi$ is part of a cold hidden sector at initial temperature $T_\varphi\simeq0$. Correspondingly, this implies that $f_\varphi\simeq0$, and the initial number density of $\varphi$ is negligible 
\beq 
n_\varphi\equiv \frac{g_\varphi}{2\pi^3} \int\dd^3 p f_\varphi\simeq0~,
\label{n}
\eeq
where $g_\varphi$ is the number of internal degrees of freedom of $\varphi$. Therefore, the latter term in the Boltzmann equation proportional to $f_\varphi$ (the back-reaction) can be neglected. This is the standard picture of the freeze-in scenario, which we shall adopt throughout. Further, we assume here that the portal operator is always sufficiently feeble that it does not bring the hidden sector into thermal equilibrium with the visible sector. We shall examine the specific requirement of this condition in Sect.~\ref{EqCo}.

  It follows that the Boltzmann equation can be rewritten as an integral with respect to centre of mass energy  as follows \cite{Hall:2009bx,1d}
\beq
 \dot{n}_\varphi+ 3Hn_\varphi  
 &\simeq \frac{3T}{512\pi^6}\int_{m_\varphi^2}^\infty \dd s\ \dd \Omega \ 
 P_{\phi\psi_1} P_{\varphi\psi_2}
 \ |\mathcal{M}|^2 _{ \phi \psi_1 \rightarrow  \varphi \psi_2} 
 \frac{1}{\sqrt{s}} ~ K_1 \left(\frac{\sqrt{s}}{T}\right) ~,
\label{2.4}
\eeq 
where $K_1$ denotes a Bessel function of the second kind and
\beq
P_{ij}=\frac{1}{2\sqrt{s}}\sqrt{s-(m_i+m_j)^2}\sqrt{s-(m_i-m_j)^2}~.
\label{Pij}
\eeq
In the limit that the particle masses involved in the scattering are negligible compared to the temperature, scattering via the dimension five operator $\frac{1}{\Lambda}\varphi\bar\psi_1\psi_2\phi$ is described by a matrix element the form
\beq
|\mathcal{M}|^2_{ \phi \psi_1 \rightarrow  \varphi \psi_2} \sim \frac{s}{\Lambda^2}~,
\label{dim5}
\eeq
where $\sqrt{s}$ is the centre of mass energy of the scattering at temperature $T$. Unless otherwise stated, throughout this paper we assume that the masses of the various states are substantially smaller than both $\Lambda$ and the reheat temperature $T_{\rm RH}$.

It follows from eq.~(\ref{2.4}) \& (\ref{dim5}) that the Boltzmann equation reduces to the form \cite{Hall:2009bx,1d}\footnote{Neglecting the mass in the lower limit of the integral leads to only percent-level deviations in the result.}
\beq
\dot{n}_\varphi+3H n_\varphi  &\simeq \frac{T}{512\pi^5 \Lambda^2} \int_0^\infty \dd s \ s^{3/2} K_1(\sqrt{s}/T)
\simeq \frac{ T^6}{16 \pi^5 \Lambda^2}~.
\label{1d}
\eeq
Using the relation $\dot T=-HT$ \cite{Kolb:1990vq}, this can be re-expressing in terms of the yield $Y\equiv\frac{n}{S}$ (where $S$ is the entropy density) to obtain
\beq
\frac{\dd Y_{\rm UV}}{\dd T}&
\simeq -\frac{1}{SHT} \frac{T^6 }{16 \pi^5 \Lambda^2} 
 \simeq - \frac{ 45 M_{\rm Pl}}{1.66 \times 32 \pi^7 g_*^S\sqrt{g_*^\rho} } \frac{ 1}{\Lambda^2}~,
\label{h1} \eeq
in terms of the effective number of degrees of freedom in the bath $g^{S,\rho}_*$ \cite{Kolb:1990vq}. Using the definitions $S=\frac{2\pi^2g_*^ST^3}{45}$ and $H=\frac{1.66\sqrt{g_*^\rho}T^2}{M_{\rm{Pl}}}$, for $M_{\rm{Pl}}$ the (non-reduced) Planck mass, then integrating with respect to temperature (between $T=0$ and $T=T_{\rm RH}$) gives \cite{Hall:2009bx} 
 \beq 
 Y_{\rm UV} &
 \simeq  \frac{180}{1.66\times(2\pi)^7 g_*^S\sqrt{g_*^\rho}} 
\left( \frac{T_{\rm RH}  M_{\rm Pl}}{ \Lambda^2 }\right)~.
\label{2-2}
\eeq
The important thing to note is that the yield depends on the reheat temperature of the visible sector. This is in contrast to the case of freeze-in via renormalisable interactions, in which the yield only depends on the coupling $\lambda$ and particle masses \cite{Hall:2009bx}. As we reproduce in Appendix \ref{ApA}, the DM yield due to IR freeze-in is parametrically 
\beq
Y_{\rm IR}\sim \lambda^2 M_{\rm Pl}/m_{\rm DM}~.
\eeq

With the above example in mind, we extend this analysis to consider a range of effective operators of varying mass dimension and involving different combinations of fields. It is important to recognise that operators of large mass dimension typically lead to many-body final states. Indeed, as we examine below, even the simplest extension of the previous example to dimension five operators $\frac{1}{\Lambda}\phi_1\phi_2\phi_3\phi_4\varphi$ involving four bath scalars $\phi_i$ and scalar DM $\varphi$, which allows freeze-in production via scattering $\phi_1\phi_2\rightarrow\phi_3\phi_4\varphi$, leads to a 3-body phase space.

The Boltzmann equation describing DM production via $2\rightarrow3$ scattering is given by
\beq
\dot{n}_\varphi+ 3Hn_\varphi = 6 \int 
\dd \Pi_1 \dd\Pi_2~f_1f_2|\mathcal{M}|^2_{2\rightarrow3} ~ \text{DLIPS}_3~,
\eeq
where $\text{DLIPS}_3$ denotes the differential Lorentz invariant phase space for 3-body final states and the numerical prefactor accounts for permutations of initial and final states. An evaluation of the 3-body phase space (see Appendix \ref{ApB}) allows the Boltzmann equation to be rewritten in a form  reminiscent of eq.~(\ref{1d})
\beq
 \dot{n}_\varphi+ 3Hn_\varphi  
 = \frac{6T}{(4\pi)^7}\int_0^\infty \dd s\   s^{3/2}\ |\mathcal{M}|_{\rm 2\rightarrow3}^2~ K_1 \left(\frac{\sqrt{s}}{T}\right) ~.
\label{text}
 \eeq
We have assumed here that the final state masses can be neglected.
By dimensional analysis the associated matrix element is parametrically
 \beq
|\mathcal{M}|^2_{2\rightarrow3} \sim \frac{1}{\Lambda^2}~.
\label{M2-3}
\eeq
Substituting this into the Boltzmann equation
and expressing our result in terms of the yield we obtain
\beq
 \frac{\dd Y_\varphi}{\dd T}  
& \simeq -\frac{1}{SHT}  \frac{6T}{(4\pi)^7}\frac{1}{\Lambda^2}\int_0^\infty \dd s\   s^{3/2} ~ K_1 \left(\frac{\sqrt{s}}{T}\right)~.
 \eeq
Performing the integrals over $s$ and, subsequently, temperature we find the form of the DM yield 
\beq
 Y_\varphi  
 \simeq \frac{135}{1.66\times (2\pi)^9g_*^S\sqrt{g_*^\rho}}
\left(\frac{T_{\rm RH}M_{\rm{Pl}}}{\Lambda^2}\right)~.
\label{r1} 
\eeq
Up to a numerical suppression of $\sim10^{-2}$, this is similar in form to eq.~(\ref{2-2}) and, notably, also depends linearly on the reheat temperature.

The DM yield may be related to the relic density as follows
\beq
\Omega_{\varphi}
=\frac{m_\varphi Y_{\varphi}  S_0}{\rho_c}
\simeq
0.2\times \left(\frac{m_\varphi}{1~{\rm TeV}}\right) \left(\frac{Y_{\varphi}}{10^{-13}}\right)~,
\label{yield}
\eeq
where $\rho_c$ denotes the critical density and $S_0$ is the present day entropy density, evaluated at $T_0=2.75~{\rm K}\sim10^{-4}~{\rm eV}$. In the latter expression we have approximated $\sqrt{g_*^\rho}g_*^S\sim10^3$. We can choose judicious parameter values such that the observed relic density ($\Omega_{\varphi}h^2\approx 0.1$) is obtained for a given value of the DM mass. For example, choosing a canonical DM mass of 1 TeV, eq.~(\ref{r1}) can be rewritten
\beq
 Y_{\varphi}&
 \simeq 10^{-13}\times \left(\frac{T_{\rm RH}}{3\times10^8 ~{\rm GeV}}\right)
 \left(\frac{10^{16}~{\rm GeV}}{\Lambda}\right)^2 ~.
\eeq

It should be noted that in non-minimal models the DM produced via freeze-in might be able to subsequently annihilate. This would introduce further terms in the Boltzmann equation. If there are additional hidden sector interactions, and light hidden sector states into which the DM can annihilate, then this can give rise to a period of annihilation and freeze-out in the hidden sector, which may dilute the DM relic density. To maintain a degree of predictability, throughout we shall assume that there are no such additional hidden sector interactions which can lead to DM pair annihilation. On the other hand, if the DM only interacts via the suppressed portal operator then, as the DM never enters thermal equilibrium, the rate of annihilation back to the visible sector is always negligible compared to the rate of production. In some sense the DM is immediately frozen-out on production.


\subsection{High dimension operators with many-body final states}
\label{s2.2}

We would like to understand how this generalises to operators of increasing mass dimension.
For UV freeze-in involving an operator of mass dimension $n+4$, the cut-off enters in the denominator of the yield as $\Lambda^{2n}$. Thus, still assuming that none of the fields involved acquire non-zero VEVs, the generic expectation is that the DM yield $Y_{(n)}$ due to this operator should scale as follows
\beq
Y_{(n)}\sim\frac{M_{\rm Pl}T^{2n-1}_{\rm RH}}{\Lambda^{2n}}~,
\label{scale}
\eeq
the factor of $M_{\rm Pl}$ coming from the Hubble parameter. One important issue which arises, however, is that the phase space becomes increasingly large and complicated. Here we shall make certain assumptions and approximations to obtain an order-of-magnitude estimate of the yield.

Consider the dimension-$(n+4)$ operator $\frac{1}{\Lambda^{n}}\phi_1\phi_2\cdots\phi_{n+3}\varphi$ which corresponds to $(n+2)$-body final state phase space  for scattering events $\phi_1\phi_2\rightarrow\phi_3\cdots\phi_{n+3}\varphi$. Variant operators might be considered but this toy example should be illustrative of a more general issue regarding the interplay between mass dimension and phase space. For simplicity here we assume that there is only a single relevant high dimension operator; the converse scenario would imply a sum over the various operators in the Boltzmann equation.

The Boltzmann equation describing DM production via $2\rightarrow n+2$ scattering is given by\footnote{We neglect here permutations of initial states. If included this leads to a combinatorial enhancement, but as the DM abundance is highly sensitive to $\Lambda$, for $n>1$ this is of lesser importance.}
\beq
\dot{n}_\varphi+ 3Hn_\varphi = \int 
\dd \Pi_1 \dd\Pi_2~f_1f_2|\mathcal{M}|^2_{(n)} ~ \text{DLIPS}_{(n+2)}~.
\eeq
The differential phase space grows like $\text{DLIPS}_{(n+2)}\propto s^n$ and we make the approximation
\beq
\text{DLIPS}_{(n+2)}\sim \Big[\frac{s}{4\pi^2}\Big]^{n}\text{DLIPS}_{(2)}~.
\eeq
The square bracket provides a parametric estimate of the additional phase space suppression due to the $(n+2)$-body final state.  This is a somewhat crude approximation, but for low $n$  should give an order-of-magnitude estimate. For increasing $n$ the suppression to the cross section coming from the phase space should become more severe and strongly suppress operators of high ($n\gg 1$) mass dimension. From dimensional analysis
\beq
|\mathcal{M}|_{(n)}^2 \sim \left(\frac{1}{\Lambda^{2}}\right)^{n}~,
\eeq
and thus the Boltzmann equation can be expressed 
\beq
\dot{n}_\varphi+3Hn_\varphi &\simeq \frac{2T}{(4\pi)^5\Lambda^{2n}}~\left[\frac{1}{4\pi^2}\right]^{n} \int_0^\infty \dd s \ s^{(2n+1)/2}~ K_1(\sqrt{s}/T)~.
\label{Mxn}
\eeq
We will check the resulting estimate against the 3-body result calculated in Sect.~\ref{S2.1}.

The integral over $s$ has a closed form expression which for $n\in\mathbb{N}$ is given by
\beq
\int_0^\infty \dd s \ s^{(2n+1)/2} K_1(\sqrt{s}/T)=4^{n+1} T^{2 n+3} n!(n+1)!~.
\eeq
Using this result the Boltzmann equation can be rewritten 
\beq
 \dot{n}_\varphi+3Hn_\varphi \simeq \frac{1}{(2\pi)^7}\left(\frac{n!(n+1)!}{\pi^{2n-2}}\right)~\frac{T^{2n+4}}{\Lambda^{2n}} ~.
\eeq
From the above we obtain an expression for the DM yield 
\beq
 Y_{(n)} &
 \simeq 
 \frac{90}{1.66\times(2\pi)^{9}\sqrt{g_*^\rho}g_*^S}~\left(\frac{n!(n+1)!}{\pi^{2n-1}}\right)
 \int_0^{T_{\rm RH}}\dd T~
 \frac{M_{\rm{Pl}}T^{2 n-2}}{\Lambda^{2n}}\\
 & \simeq 
 \frac{90}{1.66\times (2\pi)^{9} \sqrt{g_*^\rho}g_*^S} ~\frac{1}{2n-1}~\left(\frac{n!(n+1)!}{\pi^{2n-1}}\right) 
 \left(\frac{M_{\rm{Pl}}T_{\rm RH}^{2 n-1}}{\Lambda^{2n}}\right) ~.
\label{Yn}
\eeq 
The form of eq.~(\ref{Yn}) conforms with our expectations for the parametric scaling discussed in eq.~(\ref{scale}). Moreover, this shows that a range of operators of varying mass dimension should be able to reproduce the observed relic density. We consider some examples below.

Firstly, we can check this result against our 3-body calculation by setting $n=1$
\beq
 Y_{(1)}
&  \simeq 
\frac{360}{1.66\times(2\pi)^{10}}\frac{1}{\sqrt{g_*^\rho}g_*^S}\frac{M_{\rm{Pl}}T_{\rm RH}}{\Lambda^{2}}~.
\label{Y1}
\eeq 
Comparing with eq.~(\ref{r1}), we find that these two expressions agree up to an $\OO(1)$ factor. For suitable parameter values the observed relic density can be reproduced for a large range of DM masses, comparing with eq.~(\ref{yield}). As a concrete example, consider a model with 1 TeV DM, a yield of appropriate magnitude is found for
\beq
Y_{(1)}&
 \simeq 10^{-13}\times \left(\frac{T_{\rm RH}}{3\times10^8 ~{\rm GeV}}\right)
 \left(\frac{\rm 10^{16}~ GeV}{\Lambda}\right)^2 ~.
\eeq
Observe, unlike typical models of freeze-out and IR freeze-in, the yield is independent of $m_\varphi$, the DM mass, provided $m_\varphi\ll T \ll \Lambda$. Thus in UV freeze-in one can find the observed DM relic density for different values of the DM mass by simply rescaling the yield.

Taking a few further examples, consider dimension-six ($n=2$) and dimension-seven  ($n=3$) operators (with 4 and 5 body final states, respectively), the estimates for the freeze-in yield of $\varphi$ for these cases are
  \beq
Y_{(2)} & \simeq 10^{-13}
 \left(\frac{T_{\rm RH}}{5\times10^{13}~ {\rm GeV}}\right)^3
 \left(\frac{\rm 10^{16}~ GeV}{\Lambda}\right)^4
 ~,\\[5pt]
 Y_{(3)} & \simeq 10^{-13}
 \left(\frac{T_{\rm RH}}{4\times 10^{14} ~{\rm GeV}}\right)^5
 \left(\frac{\rm 10^{16}~ GeV}{\Lambda}\right)^6.
 \label{Y23}
 \eeq
Thus for a given operator and fixed UV scale, the observed DM abundance can typically be obtained by adjusting the reheat temperature. We shall comment in Sect.~\ref{s3.4} on motivated values of $T_{\rm RH}$. Moreover, we learn that in the presence of multiple high dimension portal operators of varying mass dimension (but common $\Lambda$),  generally the physics will be determined by the operator(s) with smallest mass dimension. This conforms with the standard intuition regarding effective field theory.


\subsection{VEV expansions of high dimension operators}
\label{s2.3}

If any of the fields involved in the high dimension operator acquire VEVs then, by expanding the operator around the vacuum, one can construct a sequence of terms dressed by couplings of different mass dimension \cite{Hall:2009bx}. Let us again examine the simple example given in \cite{Hall:2009bx}, involving the dimension-five operator in eq.~(\ref{JMR}). Suppose that the bath scalar field has a non-zero VEV $\langle\phi\rangle\neq0$, expanding around this VEV gives
 \beq
\mathcal{L}\supset 
\lambda\bar\psi_1\psi_2\varphi
+\frac{1}{\Lambda}\phi\bar\psi_1\psi_2\varphi~,
\label{VEV}
\eeq
where we have identified $\lambda\equiv\frac{\langle\phi\rangle}{\Lambda}$. To ensure validity of the effective field theory we will assume that $\langle\phi\rangle\ll \Lambda$ and therefore $\lambda\ll1$.
In this example the yield will receive both an IR contribution from the first term (which appears as an operator with a dimensionless coupling after symmetry breaking) and a UV contribution from the latter term. The IR contribution is assumed to be generate by decays $\psi_1\rightarrow\psi_2\varphi$, this is reproduced in eq.~(\ref{1-2}) of the Appendix. By calculating the two contributions, it can be shown  \cite{Hall:2009bx} that the yield is dominated by the IR contribution if 
\beq
\frac{3\pi^3 \langle\phi\rangle^2}{T_{\rm RH} m_{\psi_1}}>1~.
\label{cond}
\eeq
It is notable that this condition does not depend on $\Lambda$.

Let $T_*$ be the critical temperature associated to the spontaneous breaking of some symmetry, due to a scalar field involved in the UV freeze-in operator developing a VEV. It should be expected that $T_*\sim\langle\phi\rangle$. The VEV expansion of eq.~(\ref{VEV}) is only valid for $T<T_*$, but as the yield due to IR operators is temperature independent, it will not depend on the temperature $T_*$ at which the IR operator is generated. Thus it is not required that $T_{\rm RH}<T_*$ in this case. 

The situation is more complicated if the VEV expansion leads to additional UV freeze-in operators.
For $T_*<T_{\rm RH}$,   one can find UV contributions which depend on $T_*$, rather than $T_{\rm RH}$. This is because for temperatures  $T > T_*$  there is no VEV expansion until thermal evolution (due to expansion) causes $T$ to drop below $T_*$. We shall illustrate this with an example below. Importantly, if the  phase transition happens after the point at which DM freeze-in terminates, i.e.~below the mass of the bath states involved in the freeze-in process ($T_* < m_{\rm bath}$),
then no further (UV or IR) contributions will be generated.

Let us consider an example in which the VEV insertion does not lead to operators with dimensionless coefficients, but results in several UV contributions.  One manner of realising this scenario is by dressing the SM Yukawa operators with a pair of DM states
\beq
\mathcal{L}\supset \frac{1}{\Lambda^3}H\bar Qu^c\bar \chi\chi ~~ \longrightarrow~~
 \frac{v}{\Lambda^3}\bar uu\bar \chi\chi+ \frac{1}{\Lambda^3}h\bar uu\bar \chi\chi~.
\label{ex1}
\eeq
Note that the DM can be stabilised by the assumption of $\chi$-parity, such that the Lagrangian is invariant under $\chi\rightarrow-\chi$. After electroweak symmetry breaking (EWSB) one makes the appropriate expansion around the vacuum to obtain the four fermion operator. 
Henceforth we shall call $\OO_1$ the operator dressed by $v/\Lambda^3$ and the latter term $\OO_2$.

As one might anticipate the VEV expansion leads to two operators which provide UV contributions to the yield. The operator $\OO_2$ leads to freeze-in via $2\rightarrow3$ scattering processes such as $hq\rightarrow q\bar\chi\chi$. From dimensional analysis the matrix element is of the form 
\beq
|\mathcal{M}|_{\OO_2}^2\sim\frac{s^2}{\Lambda^6}~.
\eeq 
As previously, we can describe the production of DM through the Boltzmann equation given in eq.~(\ref{text}). It follows that the DM yield due to $\OO_2$ is of the form 
\beq
Y_{\OO_2}
 = \frac{9}{8\pi^9} \frac{45}{1.66\sqrt{g_*^\rho}g_*^S}
 \left(\frac{T^{5}_{\rm RH}M_{\rm{Pl}}}{\Lambda^6}\right)~.
\label{YO2}
 \eeq

Similarly, the operator $\OO_1$ results in DM production via $2\rightarrow2$ scattering $\bar qq\rightarrow \bar\chi\chi$
\beq
|\mathcal{M}|_{\OO_1}^2\sim  \langle H\rangle^2 \left(\frac{s^2}{\Lambda^6}\right)~.
\eeq 
The relevant Boltzmann equation is analogous to eq.~(\ref{1d})
\beq
\dot{n}_\varphi+3H n_\varphi  
& \simeq \frac{ \langle H\rangle^2 T}{512\pi^5 \Lambda^6} \int_0^\infty \dd s \ s^{5/2} K_1(\sqrt{s}/T)
& \simeq \frac{3  \langle H\rangle^2 T^8}{2\pi^5 \Lambda^6}
~.
\eeq
Hence the yield is given by
\beq
Y_{\OO_1}
& \simeq  \frac{3\times45}{1.66\times4\pi^7}\frac{1}{\sqrt{g_*^\rho}g_*^S}
\left(\frac{M_{\rm{Pl}}}{\Lambda^6}\right)
\int^{T_{\rm max}}_{0} \langle H\rangle^2 ~T^2~ {\rm d}T~.
\label{YO1}
\eeq
The maximum temperature $T_{\rm max}$, which is the upper limit of the integral, depends on whether the reheating temperature is above or below the critical temperature $T_*$ at which the Higgs develops a VEV and can be expressed as $T_{\rm max} = {\rm min}(T_*~,~T_{\rm RH})$.
Thus there are two possible cases, which we examine below, depending on whether the reheat temperature is above or below the phase transition. As the operator in eq.~(\ref{ex1}) involves the Higgs VEV, the critical temperature is around  $T_{*}\sim100~{\rm GeV}$, associated to the electroweak phase transition (EWPT).

Provided $T_*>T_{\rm RH}$ the VEV expansion is valid, and thus the portal operator $\OO_1$ is active, for all physically relevant temperatures. Therefore $T_{\rm max}=T_{\rm RH}$, and identifying $\langle H\rangle=v$, the ratio of the two contributions is
\beq
\frac{Y_{\OO_2}}{Y_{\OO_1}}=\frac{9}{2\pi^2}
\left(\frac{T^{5}_{\rm RH}}{T^3_{\rm max}v^2 }\right)
=\frac{9}{2\pi^2}
\left(\frac{T_{\rm RH}}{v}\right)^2~.
\eeq
For VEV expansions which do not lead to dimensionless couplings, because $T_*\sim v$ one expects  $Y_{\OO_2}/Y_{\OO_1}<1$ for $T_{*}>T_{\rm RH}$. More generally, if $T_{*}>T_{\rm RH}$ the expectation is that the contribution coming from the operator in the VEV expansion dressed by the coefficient with smallest (negative) mass dimension will dominate the yield.

For $T_{*}<T_{\rm RH}$ the phase transition takes place during the cooling of the thermal bath, in this case production via the VEV expanded operator only occurs for $T<T_*$ and the dominant contribution will be generated at $T\sim T_*$. Further, thermal fluctuations may be important in determining the field expectation value and it is expected\footnote{A more careful study of these thermal effects would be of interest, but it is beyond the scope of this work.}  that $\langle H\rangle\sim T$. Evaluating eq.~(\ref{YO1}), the ratio of the contributions coming from eq.~(\ref{ex1}) in this case is instead
\beq
\frac{Y_{\OO_2}}{Y_{\OO_1}}
\simeq\frac{15}{2\pi^2}
\left(\frac{T_{\rm RH}}{T_*}\right)^5
~.
\eeq
Since by assumption  $v<T_{\rm RH}$, the operator $\OO_2$ is typically dominant. 
More generally, we expect that for alternative operators typically the term with no explicit VEVs in the expansion around the vacuum will provide the most significant contribution to the yield.
 In the case that the VEV expansion generates a dimensionless coupling, the associated yield is independent of $T_{\rm RH}$ and $T_*$, and the criteria under which this operator dominates will be described by an equation analogous to eq.~(\ref{cond}).

For scattering processes $hq\rightarrow q\bar\chi\chi$ to occur in the thermal bath it is required that the SM states can be thermally produced. This implies that $T_{\rm RH}\gtrsim100$ GeV if $H$ is the SM Higgs. In the case that the Higgs is thermally produced, this pushes the model into the regime in which the operator $\OO_2$ always gives the dominate contribution. In the converse scenario that the Higgs is not thermally produced then  $\OO_2$ will be exponentially suppressed, but $\OO_1$ can still potentially lead to DM production.

Further complications can arise if the VEV expansion gives several terms, and thus multiple contributions to the DM yield, or if there are multiple scalar fields, especially if the scalars develop VEVs due to spontaneous symmetry breaking taking place at different scales. We shall leave these more complicated possibilities until they arise in motivated examples.


\section{Sector equilibration constraints}
\label{EqCo}

For the DM relic density to be established by freeze-in production (rather than freeze-out) it is imperative that the DM is not brought into thermal equilibrium with the visible sector due to its interactions through the portal operator. For the case of IR freeze-in via renormalisable interactions, for instance due to $\lambda\bar\psi_1\psi_2\varphi$, it has been argued \cite{Cheung:2010} that the hidden sector should not thermalise with the visible sector, provided  $\lambda\lesssim10^{-6}\sqrt{m_{\rm bath}/100~{\rm GeV}}$.\footnote{In the case that a VEV expansion gives a renormalisable operator, then identifying $\lambda=\langle \phi_1 \rangle\cdots\langle \phi_n \rangle/\Lambda^n$ one should apply this IR constraint, in addition to requirements on high dimension operators in this section.} In this section we derive an analogous condition for the case that the portal is due to a non-renormalisable operator, leading to UV freeze-in. 

It will be useful to introduce the freeze-out temperature $T_{\rm FO}$, the temperature at which a state in thermal equilibrium decouples from the thermal bath, defined such that at $T=T_{\rm FO}$
\beq
n \langle \sigma v \rangle =H ~.
\label{3.1}
\eeq
The requirement that the DM is always out of thermal equilibrium is equivalent to $Y<Y^{\rm eq}$. This implies two conditions:
\begin{itemize}
\item For $T_{\rm RH}>T>m_{\rm DM}$, the bath number density is $n^{\rm eq}\simeq T^3/\pi^2$ and thus the DM number density is  non-equilibrium provided $n_{\rm DM}\ll T^3/\pi^2$.
\item For $m_{\rm DM}>T$, the equilibrium number density is Boltzmann suppressed and thus to avoid $n_{\rm DM}$ coinciding with $n^{\rm eq}$ it is required that DM freeze-out occurs for $T_{\rm FO}\gg m_{\rm DM}$.
\end{itemize}
The UV operator freezes-in a DM abundance at $T_{\rm RH}$, which is effectively frozen-out. Clearly, an upper bound is given by the scenario in which a near thermal abundance is generated in the early universe, subsequently decouples, and then evolves to the present modified only by entropy conservation.  Including the entropy factor, this gives (for a real scalar DM particle) 
\begin{equation*}
n_{\rm DM} (T_0)=
\left(\frac{s_0}{s(T_{\rm RH})}\right)n_{\rm DM}(T_{\rm RH})=
\left(\frac{g_*^S(T_0)}{g_*^S(T_{\rm RH})}\right)\left( \frac{1.2 T_0^3}{\pi^2}\right)~.
\end{equation*}
Further, we use that the DM relic density is given by
\begin{equation*}
\Omega_{\rm DM} = \frac{n_{\rm DM} (T_0)~m_{\rm DM}}{\rho_c}~.
\end{equation*}
Comparing with the observed value $\Omega_{\rm DM} \approx 0.2$, this gives a bound on the DM mass
\begin{equation}
m_{\rm DM} ~\gtrsim~\frac{g(T_{\rm RH})}{g(T_0)}
\frac{\pi^2\rho_c\Omega_{\rm DM}}{1.2T_0^3}~\simeq~0.4~{\rm keV}~,
\label{0a}
\end{equation}
where we have used that $g(T_0)/g(T_{\rm RH}) \simeq 3.36/100$.
Thus this places a model independent bound on the DM mass.

Next we examine the second requirement: $T_{\rm FO}\gg m_{\rm DM}$. The freeze-out temperature can be found by solving eq.~(\ref{3.1}),  which we expand below
\beq
 \langle \sigma v \rangle \left(\frac{T_{\rm FO}^3}{\pi^2}\right) = H(T_{\rm FO})  \simeq1.66\sqrt{g_*^\rho}~\frac{T_{\rm FO}^2}{M_{\rm Pl}} ~.
\label{lhrh}
\eeq
Thus the requirement that freeze-out occurs before the mass threshold $m_{\rm DM}$ is given by 
\beq
T_{\rm FO}
  \simeq~\frac{1.66\sqrt{g_*^\rho}\pi^2}{M_{\rm Pl} \langle \sigma v \rangle} \gg m_{\rm DM}~.
\label{TFO}
\eeq
For a specific portal operator this can be re-expressed in terms of the UV scale $\Lambda$ at which the operator is generated. Let us consider an example. Recall the Lagrangian term studied previously for UV freeze-in by $2\rightarrow2$  scattering: $ \mathcal{L} \supset \frac{1}{\Lambda}\varphi\bar\psi_1\psi_2\phi$. In this case the matrix element is given by eq.~(\ref{dim5}) and the associated cross section is  
\beq
\langle \sigma v\rangle \sim  \left(\frac{1}{4\pi}\right)^3 \frac{1}{\Lambda^2}~.
\label{sigv1}
\eeq
It follows that the requirement eq.~(\ref{TFO}) can be expressed for this operator as below
\beq
 \Lambda \gg  \left(\frac{m_{\rm DM}M_{\rm Pl}}{1.66\times2^6\pi^5\sqrt{g_*^\rho}}\right)^{1/2}
 \simeq10^7~{\rm GeV}
 \left(\frac{m_{\rm DM}}{\rm 0.4~keV}\right)^{1/2}
~.
\eeq
Moreover, when combined with the constraint of eq.~(\ref{0a}), this implies an extreme lower bound on the scale of new physics of $10^7$ GeV, as indicated above.

%
 \begin{figure}[t!]
\begin{center}
\includegraphics[width=51mm]{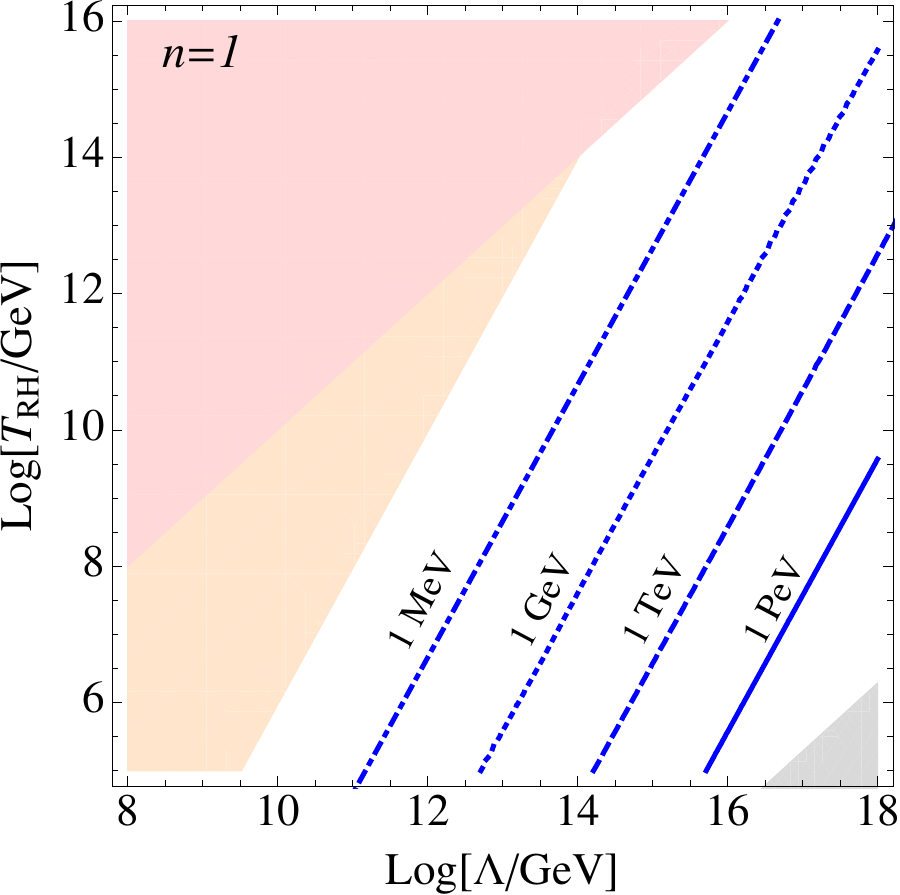}
\includegraphics[width=51mm]{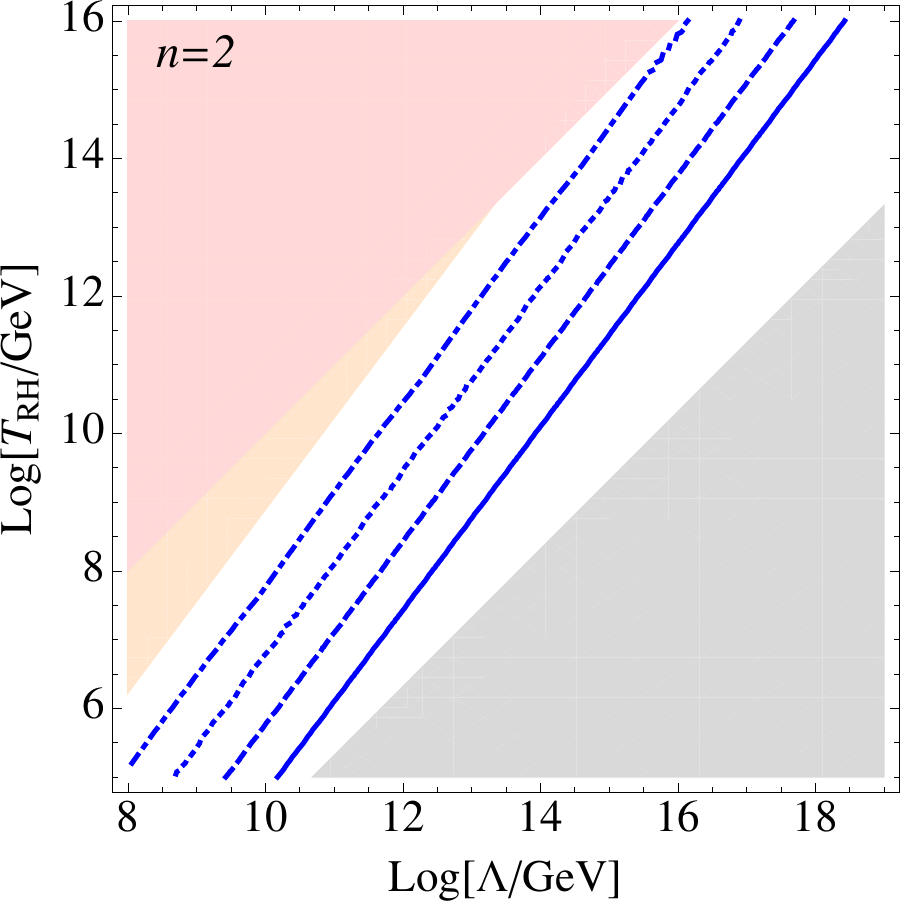}
\includegraphics[width=51mm]{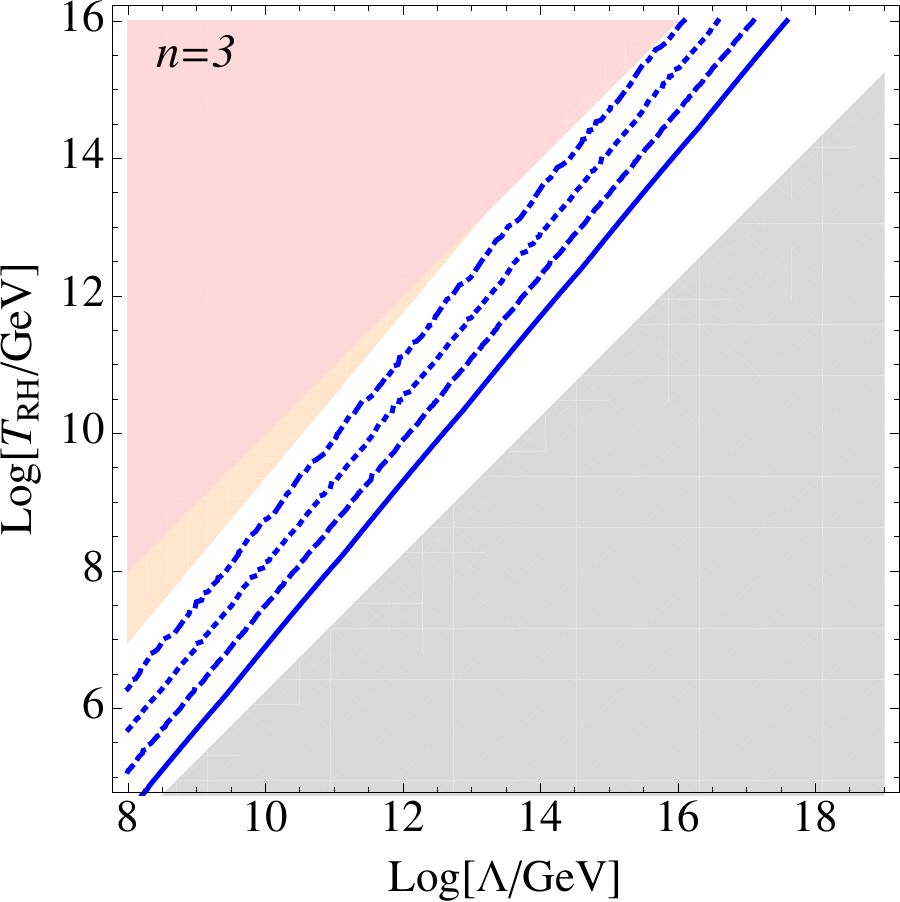}
\caption{We consider operators $\frac{1}{\Lambda^{n}}\phi_1\phi_2\cdots\phi_{n+3}\varphi$, as in Sect.~\ref{s2.2} (with $n=1,2,3$), for DM with mass $1$ PeV (solid), $1$ TeV (dashed), $1$ GeV (dotted), and 1 MeV (dot-dashed). Contours indicating values of $T_{\rm RH}$ and $\Lambda$ appropriate to match the observed DM relic density via UV freeze-in are shown, following  eq.~(\ref{r1}), (\ref{yield}) \& (\ref{Yn}).   The grey shaded areas indicate regions of parameter space in which the observed relic density can not be obtained. The parameter space highlighted in red shows where the effective theory breaks down and thus the UV freeze-in picture is not valid. The requirement $Y_{\rm DM}< Y^{\rm eq}$, that the DM does not come in to thermal equilibrium, equivalent to $m_{\rm DM}>0.4$ keV, is shown in orange. 
\label{Fig2}}
\end{center}
\end{figure}
%

In Fig.~\ref{Fig2} we display the constraints on the parameter space of various models. It particular we look at the UV freeze-in portals considered in Sect.~\ref{s2.2} \& \ref{s2.3}. Using the derived forms of the yields, we present contours of $\Lambda$ and $T_{\rm RH}$ which give the observed relic density $\Omega_{\rm DM}h^2\approx0.1$ for a range of DM masses. We overlay these contour plots with the relevant constraints. Towards the lower right of the parameter space in each plot the yields are low and require increasingly larger DM masses in order to reproduce the observed relic density. In certain regions of parameter space the DM mass which would be required is larger than the $T_{\rm RH}$ and thus such models can not give the correct DM abundance; this is indicated by the grey shaded regions in Fig.~\ref{Fig2}. The purple shaded regions indicate parameter values which lead to sector equilibration, as given in eq.~(\ref{0a}), the DM  enters thermal equilibrium $Y=Y^{\rm eq}$, thus the abundance will not be set via the freeze-in mechanism. In the red highlighted region $T_{\rm RH}>\Lambda$, the effective field theory breaks down, and the dark matter abundance is not set by UV freeze-in. Note also that because the DM is never in thermal equilibrium the unitarity bound of freeze-out DM \cite{Griest:1989wd} does not apply to freeze-in DM \cite{Hall:2009bx} and thus there is no upper bound on the DM mass.


\section{UV freeze-in and BSM physics}

\label{S3}

Having discussed a range of  possibilities in the context of toy models, we how turn to constructing explicit models based on extensions of the SM, motivated by outstanding problems. These BSM scenarios generally require additional states to be introduced at new physical scales above the weak scale.  One interesting possibility is that the operator(s) responsible for UV freeze-in are generate at this scale of new physics.  
We shall also discuss motivated BSM models which lead to high dimension operators with VEV expansions at temperatures above the weak scale.  It was remarked in \cite{Hall:2009bx} that the high dimension operator responsible for UV freeze-in might arise from GUT scale physics, here we examine some alternative scenarios.


\subsection{$Z'$ portal}
\label{s3.2}

First we consider the scenario in which the SM gauge group is extended by an additional U(1) gauge symmetry, which is broken at some high scale $\Lambda$ (for a general review see e.g.~\cite{Langacker:2008yv}) 
\beq
{\rm SU}(3)_c\times {\rm SU}(2)_L\times {\rm U}(1)_Y\times {\rm U}(1)'~.
\eeq
 If some visible sector states and the DM are both charged under this new gauge symmetry, then the associated massive gauge boson can provide a portal that links these two sectors. 
 
Additional U(1) gauge groups are a generic expectation of string theory compactifications, see 
e.g.~\cite{str,Blumenhagen:2006ci,Cvetic:1997wu}, as supported by scans of vacua of Heterotic string theory \cite{Anderson}. Further, GUTs based on $E_6$ or SO(10) can introduce extra U(1) factors from the breaking of these larger groups \cite{London:1986dk} 
 \beq 
 E_6\rightarrow {\rm SO}(10)\times {\rm U}(1)' \rightarrow {\rm SU}(5)\times U(1)'\times U(1)'' ~.
 \eeq  
 
In type IIB theories extra U(1)'s can arise from isolated branes; moreover, brane stacks are associated to symmetry groups ${\rm U}(N)\sim{\rm SU}(N)\times{\rm U}(1)$, where the U(1) factor is (pseudo)-anomalous \cite{str}.   This U(1) anomaly is cancelled via the Green-Schwarz mechanism, and as a result the $Z'$ acquires a mass near the string scale. It is interesting to note that in  type IIB theories the string scale can be lowered substantially compared to the Planck mass if the moduli are stabilised at LARGE volume \cite{Balasubramanian:2005zx}.

Alternatively, from an IR perspective, it is conceivable that a global quantum number of the SM model might be gauged. In the SM baryon number $B$ and lepton number $L$ appear as accidental symmetries and are typically broken in extensions of the SM, for instance GUTs. However, it is possible that some global quantum number of the SM may arise from an exact gauged symmetry. An appealing possibility is that the combination $B-L$ is gauged as this is anomaly free provided the spectrum includes right-handed neutrinos. If one assumes that DM is charged under $B-L$, then the U(1)${}_{B-L}$ gauge boson can provide a portal operator which connects the DM and the SM fermions, see e.g.~\cite{b-l}.
 If $B-L$ is gauged, typically it must be broken at a high scale to give masses to the neutrinos via the seesaw mechanism. Once the $Z'$ is integrated out this generates effective operators which connect the SM fermions and the DM, suppressed by the (intermediate) scale at which U(1)${}_{B-L}$ is broken. This can potentially lead to UV freeze-in for appropriate parameter choices.

More generally, suppose that the DM $\chi$, $\bar\chi$ and the SM fermions are charged under some new group U(1)$'$. It follows that the SM states can pair-annihilate and produce DM states via $\bar q q\rightarrow Z' \rightarrow \bar \chi\chi$. In the UV, as usual, interactions mediated by the (heavy) $Z'$ appear in the Lagrangian through the covariant derivative in the gauge invariant kinetic terms
\beq
\mathcal{L}\supset i\bar Q  \slashed D Q +  i\bar u \slashed D u +  i\bar \chi \slashed D \chi +\cdots~,
\eeq
where $\slashed D= \slashed\partial +iy'\slashed Z' +\cdots$, the ellipsis denote the gauge fields of the SM and $y'$ is the U(1)$'$ charge.
Once the $Z'$ is integrated out this leads to four-fermion interactions, suppressed by $\Lambda=y'_qy'_\chi/m_Z$, in the effective Lagrangian of the form
\beq
\mathcal{L}_{\rm eff}\supset \frac{1}{\Lambda^2}\bar Q\gamma_\mu Q  \bar \chi \gamma^\mu \chi +  \frac{1}{\Lambda^2}\bar u^c \gamma_\mu u^c  \bar \chi \gamma^\mu \chi +\cdots~.
\eeq
This is similar to $\OO_1$ studied in eq.~(\ref{ex1}), however the prefactor is different, as is the Lorentz structure, as here we have integrated out a Lorentz vector.  DM production will proceed fairly analogously
and the  DM yield is given by
(cf.~eq.~(\ref{Yn}))
\beq
Y_{(2)} & \simeq \frac{45}{1.66 \times 2^6\pi^{12} \sqrt{g_*^\rho}g_*^S} 
 \left(\frac{M_{\rm{Pl}}T_{\rm RH}^{3}}{\Lambda^{4}}\right) ~.
\eeq

It should be noted that the presence of an additional U(1) gauge group will generically lead to kinetic mixing via operators of the form $ F^{\mu\nu}F'_{\mu\nu}$. Such interactions can provide a renormalisable portal operator between the visible and hidden sectors. As we are primarily concerned here with UV freeze-in, we shall assume that such operators are negligible. This kinetic-mixing portal has been previously studied in the context of IR freeze-in \cite{Blennow:2013jba}. Discussion on  UV freeze-in via $Z'$ also appear in \cite{Mambrini}.


\subsection{The axion portal}
\label{s3.1}

We shall next consider a realisation of the simple toy model of UV freeze-in originally considered in \cite{Hall:2009bx}, and discussed here in eq.~(\ref{VEV}), based on the axion solution to the strong CP problem. This will provide an example in which a VEV expansion introduces additional freeze-in portals.
It is widely thought that the most viable solution to the strong CP problem is the Peccei-Quinn (PQ) mechanism, which dynamically sets the $\bar\theta$-parameter to zero \cite{Peccei:1977hh}.  Such `axion portals' have been contemplated previously within the context of DM freeze-out, e.g.~\cite{Nomura:2008ru}. We shall take a DFSZ-type model \cite{DFSZ}, where a type II two Higgs doublet is supplemented with an additional SM singlet scalar $S$ transforming under the PQ symmetry
\beq
S\rightarrow e^{2i\beta}S~,
\qquad
&H_{u}\rightarrow e^{2i\beta}H_u~,
\qquad
H_{d}\rightarrow e^{-2i\beta}H_d~.
\label{3.2}
\eeq
The SM fermions transform as $q_L\rightarrow e^{i\beta} q_L$ and $q_R\rightarrow e^{-i\beta} q_R$.  We further supplement this model with SM singlet Wely fermions $\chi$, $\bar\chi$ which transform with equal charges under the PQ symmetry.\footnote{It is useful for our purposes that the states are Dirac as a Majorana fermion $\chi'$ would necessarily be a singlet under the PQ symmetry and thus, in the absence of additional symmetries (e.g.~lepton number), this would allow the renormalisable operator $LH_u\chi'$.} The state $\chi$ is the DM candidate, it can be stabilised by a $Z_2$ $\chi$-parity. Potentially, a stabilising discrete symmetry might arise as a subgroup of the PQ symmetry. This might be considered as a `toy' setting, as we shall not confront the various naturalness problems  \cite{Volkas:1988cm,Kamionkowski:1992mf} which arise in such axion models, but it will illustrate the general principle.

In addition to the Yukawa couplings we can build the following Lorentz, gauge, and PQ invariant combinations of these fields: 
\beq
H_u^\dagger H_u, \quad H_d^\dagger H_d, \quad S^\dagger S,\quad  H_u^\dagger H_d S^2,\quad \chi \bar\chi^c~.
\label{bi}
\eeq
These field combinations allow us to construct the following Lagrangian terms involving the DM bilinear
\beq
\mathcal{L}\supset \frac{1}{\Lambda}S^\dagger S\chi\bar\chi^c +  M_\chi\chi\bar\chi^c +{\rm h.c.}+\cdots~.
\label{Lax}
\eeq
The scalar field $S$ develops a non-vanishing VEV $v_{a}$ at scale $f_{a}=2v_a\gg m_Z$, which preserves electroweak symmetry, but spontaneously breaks the PQ symmetry.

Let us assume the following mass hierarch: 
\beq
m_Z~\ll~ m_{\sigma}~ \lesssim~f_a~<~T_{\rm RH}~<~\Lambda~.
\label{hh}
\eeq
For $T _{\rm RH} > f_a$, UV freeze-in can proceed via scattering $S S^\dagger \rightarrow\chi\bar\chi$. 
The matrix element for this process is 
$|\mathcal{M}|^2\sim s/\Lambda^2$, which is similar in form to eq.~(\ref{dim5}). Thus the UV contribution to the yield is (up to $\OO(1)$ factors) as eq.~(\ref{2-2})
\beq
Y_{\rm UV}\sim  
\frac{360}{1.66\times(2\pi)^7 g_*^S\sqrt{g_*^\rho}}
\left( \frac{T_{\rm RH}  M_{\rm Pl}}{ \Lambda^2 }\right)~.
\label{gh}
\eeq
There is a similar $T_{\rm RH}$-dependent contributions to the yield from the operators $H_{i}^\dagger H_{i}  \chi \bar\chi^c$ for  ($i=u,d$) which has the same form to that given above. The correct relic density is found for $\Lambda\simeq10^9\sqrt{m_\chi T_{\rm RH}}$. A further model-dependent limit comes from the assumed mass hierarchy eq.~(\ref{hh})
\beq
m_{\chi}\simeq\frac{\Lambda^2}{10^{18}T_{\rm RH}}<\frac{\Lambda^2}{10^{18}m_\sigma}~.
\eeq
As we expect from the Lagrangian that $m_\chi\sim f_a^2/\Lambda$ and $m_\sigma\sim f_a$, this implies the following consistency constraint on the hierarchy of scales: $f_a\lesssim \Lambda/10^{6}$. The above requirements can be simultaneously satisfied with reasonable parameter values.

At energies above the EWPT, but after PQ breaking one can re-examine the operators appearing in eq.~(\ref{Lax}) following a VEV expansion around the vacuum of $S$.
The singlet field can be decomposed into radial and axial components
\beq
S=\left(f_a+ \frac{\sigma}{\sqrt{2}}\right) e^{i a/\sqrt{2}f_a}~.
\eeq
The axial field $a$ is identified with the axion.

Expanding around the VEV of $S$, the radial component $\sigma$ provides the following operators
\beq
\frac{1}{\Lambda} \chi \bar\chi^c S^\dagger S 
& \rightarrow 
 \frac{f_a^2}{\Lambda}  \chi \bar\chi^c+ \frac{f_a}{\Lambda}\sqrt{2} \sigma  \chi \bar\chi^c+
  \frac{1}{\Lambda}\frac{\sigma^2}{2}  \chi \bar\chi^c 
~.
\eeq
For $T\gtrsim m_\sigma$ the state $\sigma$ is part of the thermal bath, as it is kept in thermal contact via interactions involving products of the bilinear operators in eq.~(\ref{bi}) involving $H_u$, $H_d$ and $S$ if these have $\OO(1)$ coefficients.

Once the temperature drops below the PQ breaking scale (but whilst still above $m_\sigma$), freeze-in can proceed via direct decay of heavy $\sigma$ states to DM pairs, leading to an IR contribution to the yield. Comparing with the form of the IR freeze-in yield given in \cite{Hall:2009bx}, and reproduced in eq.~(\ref{1-2}) of Appendix \ref{ApA}, one finds
 \beq
Y_{\rm IR}\sim~\frac{135}{4\pi^3g_*^S\sqrt{g_*^\rho}}\frac{M_{\rm Pl} \Gamma_\sigma}{m_\sigma^2}
\sim
\frac{135}{4\pi^3g_*^S\sqrt{g_*^\rho}}\left(\frac{M_{\rm Pl} f_a^2}{m_\sigma\Lambda^2}\right)~,
\label{fd}
 \eeq 
where $\Gamma_\sigma\sim\frac{m_\sigma f_\sigma^2}{\Lambda^2}$ is the partial width of $\sigma\rightarrow\chi\bar\chi$. 
The condition under which the UV contribution will be dominant is 
\beq
\frac{Y_{\rm UV}}{Y_{\rm IR}}
~\sim~  
\frac{m_\sigma T_{\rm RH}}{f_a^2}
~\gtrsim~1
~.
\eeq

Note also that after EWSB there is a further VEV expansion involving the Higgs fields which leads to the portal operators of the form
\beq
\frac{1}{\Lambda}H_{i}^\dagger H_{i}  \chi \bar\chi^c & \rightarrow 
 \frac{v_i^2}{\Lambda}  \chi \bar\chi^c +  \frac{v_i}{\Lambda} \sqrt{2} h_i  \chi \bar\chi^c 
+\frac{1}{\Lambda}\frac{h_i^2}{2}  \chi \bar\chi^c. 
\eeq
This leads to IR and UV freeze-in contributions to the yield, similar to eq.~(\ref{gh}) \& (\ref{fd}), in the case that $m_\chi<m_Z$.

It would be of interest to embed the above model into a supersymmetric extension of the SM, similar to e.g.~\cite{Nomura:2008ru}. The advantages of a supersymmetric implementation is that the type II structure of the Higgs sector (required to employ the DFSZ model and to avoid constrained flavour changing processes) is an automatic consequence of holomorphy of the superpotential. In addition this might  also alleviate the naturalness problem \cite{Volkas:1988cm} associated with destabilising the weak scale, and the DM can be stabilised by $Z_2$ R-parity if it is the lightest supersymmetric particle.


\subsection{The reheat temperature}
\label{s3.4}

In Sect.~\ref{s3.2} \& \ref{s3.1} we have discussed motivated scales of new physics which might generate the high dimension operators. An interesting alternative to this approach is to consider special values for the reheat temperature $T_{\rm RH}$ and use this, in conjunction with the DM relic density, to identify the unknown UV scale. One drawback of this scenario is that very little is known about the reheat temperature. Precision measurements of primordial elements due to Big Bang nucleosynthesis are thought to imply that $T_{\rm RH}\gtrsim$ few MeV.  Models of inflation typically suggest an upper bound around  $T_{\rm RH}\lesssim10^{16}~{\rm GeV}$, see e.g.~\cite{Linde:2005ht}. Moreover, if $T_{\rm RH}$ is high then in principle this can lead to problems with long-lived exotic relics which can over-close the Universe, the classic example being the cosmological gravitino problem of supergravity \cite{Moroi:1993mb}. This implies the upper bound on the reheat temperature in models of supergravity is typically $T_{\rm RH}\lesssim10^{10}~{\rm GeV}$, with substantially stronger bounds if the gravitino is light.\footnote{It has been further argued \cite{Cheung:2011mg} that, with some assumptions,  once the contribution from axinos is also included this can lead to an even more stringent upper bound: $T_{\rm RH}\lesssim10^{5}~{\rm GeV}$.}

The following extreme cases may be of particular interest: 
\begin{itemize}
\item
 The maximum expected from simple models of inflation: $T_{\rm RH}\sim10^{16}~{\rm GeV}$.
 \vspace{-2mm}
\item
The upper bound from the cosmological gravitino problem:  $T_{\rm RH}\sim10^{10}~{\rm GeV}$. \vspace{-2mm}
\item
The lower bound from precision measurement of primordial elements:  $T_{\rm RH}\sim10~{\rm MeV}$.
  \end{itemize}
The first two scenarios might be motivated through considerations of environmental selection if there is some anthropic pressure which favours high $T_{\rm RH}$, with the cosmological gravitino problem imposing a catastrophic boundary at $T_{\rm RH}\sim10^{10}~{\rm GeV}$ in supersymmetric models.

Let us consider a specific example involving dimension-$(n+4)$ operator $\frac{1}{\Lambda^{n}}\phi_1\phi_2\cdots\phi_{n+3}\varphi$ of the scalar toy model studied in Sect.~\ref{s2.2}.
For DM with mass around 100 GeV the yield required to obtain the observed relic density is $Y\sim4\times10^{-12}$, as discussed in eq.~(\ref{yield}). To obtain the correct relic abundance via freeze-in through the dimension-five operator with a reheat temperature of  $T_{\rm RH}\sim10^{16}~{\rm GeV}$, requires a UV scale as indicated below 
\beq
 Y_{(1)}
 &\simeq 4\times10^{-12}
 \left(\frac{T_{\rm RH}}{ 10^{16}~{\rm GeV}}\right)
 \left(\frac{M_{\rm Pl}}{\Lambda}\right)^2~.
 \eeq
The magnitude of the UV scale $\Lambda$ in this example may be suggestive of a connection with Planck Scale physics.

It is not implausible that future observations might indicate the reheat temperature (given some assumptions regarding the model of inflation). Ultimately to test UV freeze-in DM, and disambiguate it from other frameworks, it will be necessary to determine the DM mass, the UV scale $\Lambda$ and $T_{\rm RH}$. Recently the BICEP2 collaboration claimed they had observed primordial tensor modes \cite{Ade:2014xna}, and thus could infer $T_{\rm RH}$, however this result is currently disputed \cite{Flauger:2014qra}. If the BICEP2 signal survives further scrutiny we shall comment on this in a dedicated paper.


\section{Conclusion}
\label{S4}

This work has provided an exploratory study of the model building opportunities which arise for the UV freeze-in mechanism. We considered general aspects of this scenario in the context of toy models and demonstrated that interesting and phenomenologically viable models can be constructed in motivated settings of BSM physics. In Sect.~\ref{S2}, we examined various toy models which encapsulate the fundamental features of UV freeze-in. Typically high dimension operators lead to many body final states, and the case of DM production via $2\rightarrow3$ UV freeze-in was carefully studied. Subsequently, we attempted to quantify  DM production associated to more complicated phase spaces. Further, we discussed the potential impact of spontaneous symmetry breaking on UV freeze-in, in particular, we identified a new case of interest in which the VEV expansion leads only to additional UV contributions, and does not generate an IR freeze-in portal. In this scenario the DM yield depends on both the reheat temperature and the critical temperature of symmetry breaking.  Sect.~\ref{EqCo} examined the constraints on UV freeze-in from the requirement that the hidden sector and visible sector do not equilibrate and we argued that this can lead to bounds on the DM mass.

In Sect.~\ref{S3} we presented realistic models of UV freeze-in and related these to interesting BSM scenarios. We suggested that UV freeze-in might be connected to motivated solutions of prominent puzzles of the SM, specifically we consider an example involving the Peccei-Quinn mechanism. A further example was presented in which the UV freeze-in portal is generated by integrating out a heavy $Z'$. This is appealing as $Z'$ arise in many extensions of the SM and additional U(1) gauge groups are common in realistic string compactifications. It should be evident from our discussions that UV freeze-in  offers a large range of possibilities for DM model building and that there are many interesting aspects yet to be explored.

UV freeze-in presents a new manner of obtaining non-thermal DM, with a relic abundance directly related to the reheat temperature, and provides an interesting alternative to the conventional ideas regarding the DM thermal history. For the DM relic density to be set through UV freeze-in it is required that reheating of the hidden sector is negligible and that the DM is connected to the visible sector via non-renormalisable operators. Once one assumes that the abundance of DM is initially depleted, one might argue that freeze-in via high dimension contact operators presents a more generic mechanism than IR freeze-in portals, which require very small renormalisable couplings or complicated effective operators involving several scalar fields with non-vanishing VEVs. Moreover, from a UV perspective, it is a fairly general expectation that distinct sectors in the low energy theory may become coupled through the high scale physics, and we have presented some examples of this principle in the above.


\vspace{3mm}

{\bf Acknowledgements}~
We would like to thank Joe Bramante, John March-Russell, Adam Martin, and Bibhushan Shakya for useful discussions. Also, we are grateful to the JHEP referee for their insightful comments.  This research was supported by the National Science Foundation under Grant No.~PHY-1215979. JU is grateful for the hospitality of the Centre for Future High Energy Physics, Beijing, where some of this work was undertaken.


\appendix

\section{IR freeze-in of dark matter}
\label{ApA}

The IR yield due to $2\rightarrow2$ scattering via a four-point scalar interaction with the matrix element $ |\mathcal{M}|^2=\lambda^2$ was calculated in \cite{Hall:2009bx}. This result is referenced in the text, so we give it here for completeness. The abundance of the DM $\varphi$ is initially zero, and it is produced via the operator $\lambda \varphi \phi_1\phi_2\phi_3$, where $\lambda$ is a feeble dimensionless coupling. Consider $2 \rightarrow 2$ scattering where the momenta of the incoming bath particles are labelled $p_1,p_2$ and outgoing state momenta labelled $p_3,p_\varphi$. 

The matrix element associated to scattering via this four-point interaction is $ |\mathcal{M}|^2=\lambda^2$. The Boltzmann equation which describes DM production in this set-up is 
\beq
\dot{n}_\varphi+ 3Hn_\varphi \simeq 3 \int \dd \Pi_1 \dd \Pi_2 \dd \Pi_3 \dd \Pi_\varphi f_1 f_2
|\mathcal{M}|^2 (2\pi)^4\delta^{(4)}(p_1+p_2-p_3-p_\varphi)~.
\eeq
As previously, we re-express this as an integral with respect to centre of mass energy
\beq
 \dot{n}_\varphi+ 3Hn_\varphi  
 &\simeq \frac{3T}{512\pi^6}\int_{m_\varphi^2}^\infty \dd s\ \dd \Omega \ P_{12} P_{3\varphi} \ |\mathcal{M}|^2 
 \frac{1}{\sqrt{s}} ~ K_1 \left(\frac{\sqrt{s}}{T}\right) ~,
\eeq 
where $P_{ij}$ is defined in eq.~(\ref{Pij}).
If the bath state masses can be neglected,  this reduces to
\beq
 \dot{n}_\varphi+ 3Hn_\varphi  
 &\simeq \frac{3T\lambda^2}{512\pi^5}\int_{m_\varphi^2}^\infty \dd s\  \left( \frac{s-m_\varphi^2}{\sqrt{s}} \right) K_1 \left(\frac{\sqrt{s}}{T}\right) 
~ \simeq ~
 \frac{3m_\varphi T^3\lambda^2}{128\pi^5} K_1 \left(\frac{m_\varphi}{T}\right) 
 ~.
\eeq 
Converting this to a yield one obtains the result \cite{Hall:2009bx}
\beq
Y_{\varphi}\simeq\frac{135}{512\pi^6 (1.66)g_*^S\sqrt{g_*^\rho}}\frac{M_{\rm Pl}\lambda^2}{m_\varphi}~.
\label{IR}
\eeq

For the case of direct freeze-in due to decays of heavy bath states $\psi_H$ to a lighter state $\psi_L$ in the thermal bath and DM $\varphi$ via the interaction $\lambda \bar\psi_H\psi_L\varphi$ this is instead
\beq
\dot n_\varphi + 3Hn_\varphi 
&~\simeq~ 2\int \dd \Pi_{H}\Gamma_{H}m_{H}f_{H}\\
&~\simeq~ 2\int^\infty_{m_{H}} \frac{\Gamma_{H}m_{H}}{2\pi^2} \sqrt{E_{H}-m_{H}} ~e^{-\frac{E_{H}}{T}}~\dd E_{H}\\
&~\simeq~ \left(\frac{\Gamma_{H}m_{H}^2T}{2\pi^2}\right) K_1\left(\frac{E_{H}}{T}\right)~.
\eeq
It follows that the yield is of the form  \cite{Hall:2009bx}
\beq
Y_\varphi ~\simeq~\frac{135}{8\pi^3(1.66)g_*^S\sqrt{g_*^\rho}} 
\left(\frac{M_{\rm Pl} \Gamma_H}{m_H^2}\right)~.
\label{1-2}
\eeq


\section{UV freeze-in  of dark matter via $2\rightarrow3$ scattering}
\label{ApB}

Consider $2 \rightarrow 3$ scattering where the momenta of the incoming particles are labelled $p_1,p_2$ and outgoing state momenta labelled $p_3,p_4,p_\varphi$, where $\varphi$ indicates that it associated to the DM and the other states are part of the thermal bath. The Boltzmann equation for the production of $\varphi$ via these scatterings are given by
\beq
\dot{n}_\varphi+ 3Hn_\varphi = \int \dd \Pi_1 \dd \Pi_2 f_1 f_2
|\mathcal{M}|^2 \text{DLIPS}_3~,
\eeq
where DLIPS$_3$ denotes the Differential Lorentz Invariant Phase Space for 3-body final states
\beq
{\rm DLIPS}_3 =
\dd\Pi_3\dd\Pi_4\dd\Pi_\varphi (2\pi)^4 \delta^{(4)} (p_1+p_2- p_3 - p_4 - p_\varphi)~.
\label{dlips}
\eeq

To evaluate the Boltzmann equation we shall first look at simplifying the form of the RHS; we start by observing that 
\beq
\dd^3 p_1 \dd^3 p_2 = (4 \pi |p_1| E_1 ~\dd E_1)(4 \pi |p_2|E_2 ~\dd E_2 )\frac{1}{2} \cos \theta~.
\eeq
It is convenient to make the following change of variables  (following broadly \cite{1d})
\beq
E_+ \equiv E_1 + E_2~,
\qquad E_-\equiv E_1 - E_2~, 
\qquad s = 2E_1E_2-2 |p_1| p_2|\cos \theta~.
\eeq
It follows that the volume element can be rewritten in these new variables as follows
\beq
\int\dd \Pi_1\dd \Pi_2
=\int \frac{1}{(2\pi)^4} \frac{\dd E_+ \dd E_- \dd s}{8}
=\int \frac{1}{(2\pi)^4} \frac{\sqrt{E_+^2-s}}{4}~\dd E_+ \dd s ~,
\eeq
since $|E_-| \leq \sqrt{E_+^2-s}$ the integral over $E_-$ can be evaluated $\int \dd E_- = 2  \sqrt{E_+^2-s}$. The Boltzmann equation reduces to the form
\beq
 \dot{n}_\varphi+ 3Hn_\varphi  &= \int_0^\infty \dd s \int_{\sqrt{s}}^\infty  \dd E_+  
 e^{- E_+/T} \frac{1}{(2\pi)^4} \frac{\sqrt{E_+^2-s}}{4} |\mathcal{M}|^2 \text{DLIPS}_3
 \eeq
 \beq
 &=\frac{T}{(2 \pi)^4} \int_0^\infty \dd s  \frac{\sqrt{s}}{4} |\mathcal{M}|^2 K_1 \left(\frac{\sqrt{s}}{T}\right)  \text{DLIPS}_3~.
\label{BE2}
\eeq
Now turning to the DLIPS factor; in the centre of mass frame, we have $\vec{p}_1 + \vec{p}_2= 0$ and hence by conservation of momentum $\vec{p}_3 = - (\vec{p}_4+ \vec{p}_\varphi)$. From which it follows that
\beq
E_3^2 = | p_4|^2+ |p_\varphi|^2 + 2 |p_4| |p_\varphi| \cos \theta_{4\varphi} ~,
 \eeq
 using $\vec p_4 \cdot \vec p_\varphi = |p_4| |p_\varphi| \cos \theta_{4\varphi}$. The phase space differential, given in eq.~(\ref{dlips}), reduces to
\beq
{\rm DLIPS}_3  &=  \frac{1}{(2\pi)^5} \frac{\dd^3 p_4 \ \dd^3 p_\varphi}{8 |p_\varphi| |p_4|E_3} ~ \delta (\sqrt{s}- \sqrt{| p_4|^2+ |p_\varphi|^2 + 2 |p_4| |p_\varphi| \cos \theta_{4\varphi}}-|p_4|-|p_\varphi|)
\eeq
where we have assumed that the masses of the particles are negligible compared to $\sqrt{s}$. Further, we define 
\beq
\overline{\cos \theta} \equiv \frac{s-2\sqrt{s} (|p_4|+|p_\varphi| )+2 |p_4||p_\varphi|}{2 |p_4| |p_\varphi|}~,
\eeq
such that $\overline{\cos \theta}$ is a solution to the delta function. Therefore, we can write 
\beq
\delta (\sqrt{s}- \sqrt{| p_4|^2+ |p_\varphi|^2 + 2 |p_4| |p_\varphi| \cos \theta_{4\varphi}}-|p_4| -|p_\varphi|) = \frac{\delta(\cos \theta_{4\varphi} -\overline{\cos \theta})}{|p_4| |p_\varphi|/E_3}~.
\eeq
It follows, after some simplifications, that 
\beq 
{\rm DLIPS}_3 = \frac{1}{(2\pi)^5} \frac{\dd \cos \theta_{4\varphi}~ \dd\phi_{4\varphi}~\dd p_4^0 ~\dd\Omega ~\dd p_\varphi^0 }{8 } ~\delta(\cos \theta_{4\varphi} - \overline{\cos \theta})~,
\eeq
where we have used that  $\dd^3p_4 = |p_4|^2 \dd(\cos \theta_{4\varphi} ) \dd \phi_{4\varphi}\dd p_4^0$ and $\dd^3 p_\varphi = |p_\varphi|^2 \dd \Omega \dd p_\varphi^0$.
As the amplitude does not depend on any angle, the angular integrals are trivial
\beq
 \int \dd\Omega = 4\pi ~, \qquad
  \int \dd \phi_{4\varphi} = 2 \pi~, \qquad
   \int \dd\cos \theta_{4\varphi} \delta (\cos\theta_{4\varphi} - \overline{\cos \theta}) =1~.
   \eeq
    Consequently, the differential phase space factor simplifies substantially 
\beq
 {\rm DLIPS}_3 = \frac{1}{(2\pi)^3} \frac{\dd p^0_4 \dd p^0_\varphi}{4}= \frac{1}{(2\pi)^3} \frac{s}{16}\dd x_1 \dd x_2 ~,
 \label{dlips3}
 \eeq
where we have made the following change of variables (following e.g.~\cite{Ciafaloni:2011sa}) in the latter equality
\beq
p_3^0 &= (1-x_1+x_2) \frac{\sqrt{s}}{2}~, \qquad
p_4^0 &= x_1 \frac{\sqrt{s}}{2}~, \qquad
p_\varphi^0&= (1-x_2)\frac{ \sqrt{s}}{2}~.
\eeq
Substituting eq.~(\ref{dlips3}) into eq.~(\ref{BE2}), and integrating over the $x_i$, we obtain the result
\beq
 \dot{n}_\varphi+ 3Hn_\varphi  &= \frac{T}{2^{13} \pi^7}\int_0^\infty \dd s  \int_0^1 \dd x_2 \int_{x_2}^1 \dd x_1 ~ s^{3/2}  |\mathcal{M}|^2 K_1 \left(\frac{\sqrt{s}}{T}\right) \\
 &= \frac{T}{(4\pi)^7}\int_0^\infty \dd s\   s^{3/2}\ |\mathcal{M}|^2 K_1 \left(\frac{\sqrt{s}}{T}\right) ~.
\eeq



\begin{thebibliography}{}   



\bibitem{Hall:2009bx}
  L.~J.~Hall, K.~Jedamzik, J.~March-Russell and S.~M.~West,
  {\em Freeze-In Production of FIMP Dark Matter,}
  JHEP {\bf 1003} (2010) 080
  [0911.1120].

\bibitem{Cheung:2010}
  C.~Cheung, G.~Elor, L.~J.~Hall and P.~Kumar,
  {\em Origins of Hidden Sector Dark Matter I: Cosmology,}
  JHEP {\bf 1103} (2011) 042
  [1010.0022].
  
    \bibitem{FI}
  C.~Cheung, G.~Elor, L.~J.~Hall and P.~Kumar,
{\em Origins of Hidden Sector Dark Matter II: Collider Physics,}
  JHEP {\bf 1103} (2011) 085
  [1010.0024].
%
  %
  L.~J.~Hall, J.~March-Russell and S.~M.~West,
  {\em A Unified Theory of Matter Genesis: Asymmetric Freeze-In,}
  [1010.0245].
%
  A.~Hook,
  {\em Unitarity constraints on asymmetric freeze-in,}
  Phys.\ Rev.\ D {\bf 84} (2011) 055003
  [1105.3728].
  %
  C.~E.~Yaguna,
  {\em The Singlet Scalar as FIMP Dark Matter,}
  JHEP {\bf 1108} (2011) 060
  [1105.1654].
  %
  C.~E.~Yaguna,
  {\em An intermediate framework between WIMP, FIMP, and EWIP dark matter,}
  JCAP {\bf 1202} (2012) 006
  [1111.6831].
  %
  X.~Chu, T.~Hambye and M.~H.~G.~Tytgat,
  {\em The Four Basic Ways of Creating Dark Matter Through a Portal,}
  JCAP {\bf 1205} (2012) 034
  [1112.0493].
  %
  P.~S.~Bhupal Dev, A.~Mazumdar and S.~Qutub,
  {\em Constraining Non-thermal and Thermal properties of Dark Matter,}
  Physics {\bf 2} (2014) 26
  [1311.5297].
  %
  J.~Unwin,
  {\em Towards Cogenesis via Asymmetric Freeze-in: The $\chi$ Who Came-in from the Cold,}
  [1406.3027].
  
 \bibitem{Blennow:2013jba}
  M.~Blennow, E.~Fernandez-Martinez and B.~Zaldivar,
  {\em Freeze-in through portals,}
  [1309.7348].

\bibitem{Kolda:2014ppa}
  C.~Kolda and J.~Unwin,
  {\em X-ray lines from R-parity violating decays of keV sparticles,}
  Phys.\ Rev.\ D {\bf 90} (2014) 023535
  [1403.5580].
  
  
\bibitem{Mambrini}
  Y.~Mambrini, K.~A.~Olive, J.~Quevillon and B.~Zaldivar,
  {\em Gauge Coupling Unification and Nonequilibrium Thermal Dark Matter,}
  Phys.\ Rev.\ Lett.\  {\bf 110} (2013) 24,  241306
  [1302.4438 ].
  %
  X.~Chu, Y.~Mambrini, J.~Quevillon and B.~Zaldivar,
  {\em Thermal and non-thermal production of dark matter via Z'-portal(s),}
  JCAP {\bf 1401} (2014) 034
  [1306.4677].
  
  \bibitem{Asaka}
  T.~Asaka, K.~Ishiwata and T.~Moroi,
  {\em Right-handed sneutrino as cold dark matter,}
  Phys.\ Rev.\ D {\bf 73}, 051301 (2006)
  [hep-ph/0512118];
%
  {\em Right-handed sneutrino as cold dark matter of the universe,}
  Phys.\ Rev.\ D {\bf 75}, 065001 (2007)
  [hep-ph/0612211].
  
\bibitem{axion}
  E.~J.~Chun, H.~B.~Kim and D.~H.~Lyth,
  {\em Cosmological constraints on a Peccei-Quinn flatino as the lightest supersymmetric particle,}
  Phys.\ Rev.\ D {\bf 62} (2000) 125001
  [hep-ph/0008139].
  %
  L.~Covi, H.~B.~Kim, J.~E.~Kim and L.~Roszkowski,
  {\em Axinos as dark matter,}
  JHEP {\bf 0105} (2001) 033
  [hep-ph/0101009].
  %
  K.~J.~Bae, K.~Choi and S.~H.~Im,
  {\em Effective Interactions of Axion Supermultiplet and Thermal Production of Axino Dark Matter,}
  JHEP {\bf 1108} (2011) 065 
  [1106.2452].
  
  \bibitem{3/2}
    J.~R.~Ellis, J.~E.~Kim and D.~V.~Nanopoulos,
  {\em Cosmological Gravitino Regeneration and Decay,}
  Phys.\ Lett.\ B {\bf 145} (1984) 181.
  %
  G.~F.~Giudice, A.~Riotto and I.~Tkachev,
  {\em Thermal and nonthermal production of gravitinos in the early universe,}
  JHEP {\bf 9911} (1999) 036
  [hep-ph/9911302].
  %
  R.~Kallosh, L.~Kofman, A.~D.~Linde and A.~Van Proeyen,
  {\em Gravitino production after inflation,}
  Phys.\ Rev.\ D {\bf 61} (2000) 103503
  [hep-th/9907124].
  %
   K.~Choi, K.~Hwang, H.~B.~Kim and T.~Lee,
  {\em Cosmological gravitino production in gauge mediated supersymmetry breaking models,}
  Phys.\ Lett.\ B {\bf 467} (1999) 211
  [hep-ph/9902291].
  
\bibitem{McDonald:2001vt}
  J.~McDonald,
  {\em Thermally generated gauge singlet scalars as self-interacting dark matter,}
  Phys.\ Rev.\ Lett.\  {\bf 88} (2002) 091304
  [hep-ph/0106249].

\bibitem{McDonald:2008ua}
  J.~McDonald and N.~Sahu,
  {\em keV Warm Dark Matter via the Supersymmetric Higgs Portal,}
  Phys.\ Rev.\ D {\bf 79} (2009) 103523
  [0809.0247].

\bibitem{Harling:2008px}
  B.~von Harling and A.~Hebecker,
  {\em Sequestered Dark Matter,}
  JHEP {\bf 0805} (2008) 031
  [0801.4015].

  \bibitem{Chung:1998zb}
  D.~J.~H.~Chung, E.~W.~Kolb and A.~Riotto,
  {\em Superheavy dark matter,}
  Phys.\ Rev.\ D {\bf 59} (1998) 023501
  [hep-ph/9802238];
%
  {\em Production of massive particles during reheating,}
  Phys.\ Rev.\ D {\bf 60} (1999) 063504
  [hep-ph/9809453].

\bibitem{Kolb:1990vq}
  E.~W.~Kolb and M.~S.~Turner,
 {\em The Early Universe,}
  Front.\ Phys.\  {\bf 69} (1990) 1.
  
\bibitem{1d}
  J.~Edsjo and P.~Gondolo,
  {\em Neutralino relic density including coannihilations,}
  Phys.\ Rev.\ D {\bf 56} (1997) 1879
  [hep-ph/9704361].
%
  P.~Gondolo and G.~Gelmini,
  {\em Cosmic abundances of stable particles: Improved analysis,}
  Nucl.\ Phys.\ B {\bf 360} (1991) 145.
  
\bibitem{Griest:1989wd}
  K.~Griest and M.~Kamionkowski,
 {\em Unitarity Limits on the Mass and Radius of Dark Matter Particles,}
  Phys.\ Rev.\ Lett.\  {\bf 64} (1990) 615.

\bibitem{Langacker:2008yv}
  P.~Langacker,
  {\em The Physics of Heavy $Z^\prime$ Gauge Bosons,}
  Rev.\ Mod.\ Phys.\  {\bf 81} (2009) 1199
  [0801.1345].
  
\bibitem{str}
  R.~Blumenhagen, M.~Cvetic, P.~Langacker and G.~Shiu,
{\em Toward realistic intersecting D-brane models,}
  Ann.\ Rev.\ Nucl.\ Part.\ Sci.\  {\bf 55} (2005) 71
  [hep-th/0502005].
  %
  E.~Kiritsis,
{\em D-branes in standard model building, gravity and cosmology,}
  Phys.\ Rept.\  {\bf 421} (2005) 105
  [Erratum-ibid.\  {\bf 429} (2006) 121]
   [Fortsch.\ Phys.\  {\bf 52} (2004) 200]
  [hep-th/0310001].

\bibitem{Blumenhagen:2006ci}
  R.~Blumenhagen, B.~Kors, D.~Lust and S.~Stieberger,
{\em Four-dimensional String Compactifications with D-Branes, Orientifolds and Fluxes,}
  Phys.\ Rept.\  {\bf 445} (2007) 1
  [hep-th/0610327].
  
\bibitem{Cvetic:1997wu}
  M.~Cvetic and P.~Langacker,
  {\em Z-prime physics and supersymmetry,}
  Adv.\ Ser.\ Direct.\ High Energy Phys.\  {\bf 21} (2010) 325
  [hep-ph/9707451].
  
\bibitem{Anderson}
  L.~B.~Anderson, J.~Gray, A.~Lukas and E.~Palti,
  {\em Two Hundred Heterotic Standard Models on Smooth Calabi-Yau Threefolds,}
  Phys.\ Rev.\ D {\bf 84} (2011) 106005
  [1106.4804].
  %
  L.~B.~Anderson, A.~Constantin, J.~Gray, A.~Lukas and E.~Palti,
  {\em A Comprehensive Scan for Heterotic SU(5) GUT models,}
  JHEP {\bf 1401} (2014) 047
  [1307.4787].

\bibitem{London:1986dk}
  D.~London and J.~L.~Rosner,
{\em Extra Gauge Bosons in $E_6$,}
  Phys.\ Rev.\ D {\bf 34} (1986) 1530.

\bibitem{Balasubramanian:2005zx}
  V.~Balasubramanian, P.~Berglund, J.~P.~Conlon and F.~Quevedo,
  {\em Systematics of moduli stabilisation in Calabi-Yau flux compactifications,}
  JHEP {\bf 0503} (2005) 007
  [hep-th/0502058].
 
  \bibitem{b-l}
  P.~Fileviez Perez and M.~B.~Wise,
    {\em Baryon and lepton number as local gauge symmetries,}
  Phys.\ Rev.\ D {\bf 82} (2010) 011901
  [Erratum-ibid.\ D {\bf 82} (2010) 079901]
  [1002.1754].
  %
  T.~R.~Dulaney, P.~Fileviez Perez and M.~B.~Wise,
    {\em Dark Matter, Baryon Asymmetry, and Spontaneous B and L Breaking,}
  Phys.\ Rev.\ D {\bf 83} (2011) 023520
  [1005.0617].
  %
  M.~Ibe, S.~Matsumoto and T.~T.~Yanagida,
  {\em The GeV-scale dark matter with B-L asymmetry,}
  Phys.\ Lett.\ B {\bf 708} (2012) 112
  [1110.5452].
  
\bibitem{Peccei:1977hh}
  R.~D.~Peccei and H.~R.~Quinn,
  {\em CP Conservation in the Presence of Instantons,}
  Phys.\ Rev.\ Lett.\  {\bf 38} (1977) 1440.
  
\bibitem{Nomura:2008ru}
  Y.~Nomura and J.~Thaler,
  {\em Dark Matter through the Axion Portal,}
  Phys.\ Rev.\ D {\bf 79} (2009) 075008
  [0810.5397].

\bibitem{DFSZ}
  M.~Dine, W.~Fischler and M.~Srednicki,
  {\em A Simple Solution to the Strong CP Problem with a Harmless Axion,}
  Phys.\ Lett.\ B {\bf 104} (1981) 199.
  %
  A.~R.~Zhitnitsky,
  {\em On Possible Suppression of the Axion Hadron Interactions.}
  Sov.\ J.\ Nucl.\ Phys.\  {\bf 31} (1980) 260
   [Yad.\ Fiz.\  {\bf 31} (1980) 497].

\bibitem{Volkas:1988cm} 
  R.~R.~Volkas, A.~J.~Davies and G.~C.~Joshi,
  {\em Naturalness Of The Invisible Axion Model,}
  Phys.\ Lett.\ B {\bf 215}, 133 (1988).

\bibitem{Kamionkowski:1992mf}
  M.~Kamionkowski and J.~March-Russell,
  {\em Planck scale physics and the Peccei-Quinn mechanism,}
  Phys.\ Lett.\ B {\bf 282} (1992) 137
  [hep-th/9202003].

    \bibitem{Linde:2005ht}
  A.~D.~Linde,
  {\em Particle physics and inflationary cosmology,}
  Contemp.\ Concepts Phys.\  {\bf 5} (1990) 1
  [hep-th/0503203].
   
\bibitem{Moroi:1993mb}
  T.~Moroi, H.~Murayama and M.~Yamaguchi,
  {\em Cosmological constraints on the light stable gravitino,}
  Phys.\ Lett.\ B {\bf 303} (1993) 289.
 
\bibitem{Cheung:2011mg}
  C.~Cheung, G.~Elor and L.~J.~Hall,
  {\em The Cosmological Axino Problem,}
  Phys.\ Rev.\ D {\bf 85} (2012) 015008
  [1104.0692].
    
\bibitem{Ade:2014xna}
  P.~A.~R.~Ade {\it et al.}  {\bf [BICEP2 Collaboration]},
  {\em Detection of B-Mode Polarization at Degree Angular Scales by BICEP2,}
  Phys.\ Rev.\ Lett.\  {\bf 112} (2014) 241101
  [1403.3985].
  
\bibitem{Flauger:2014qra}
  R.~Flauger, J.~C.~Hill and D.~N.~Spergel,
  {\em Toward an Understanding of Foreground Emission in the BICEP2 Region,}
   JCAP {\bf 1408} (2014) 039 
   [1405.7351].
  
\bibitem{Ciafaloni:2011sa}
  P.~Ciafaloni,  M.~Cirelli, D.~Comelli, A.~De Simone, A.~Riotto and A.~Urbano,
  {\em On the Importance of Electroweak Corrections for Majorana Dark Matter Indirect Detection,}
  JCAP {\bf 1106} (2011) 018
  [1104.2996].
    

\end{thebibliography}
\end{document}